\documentclass[12pt,twoside,dina4p]{article}
\usepackage{amssymb}
\def\bbl{\mathbb}
\def\got{\mathfrak}
\def\be{\vskip 0.5cm \noindent $$}
\def\ee{$$\vskip 0.5cm\noindent}

\def\Ag{{\got A}}
\def\Bg{{\got B}}

\def\Mg{{\got M}}
\def\Ng{{\got N}}

\def\sg{{\got s}}

\def\Cl{{\bbl C}}

\def\Rl{{\bbl R}}

\def\Ac{{\cal A}}
\def\Bc{{\cal B}}

\def\Hc{{\cal H}}

\def\Kc{{\cal K}}

\def\Oc{{\cal O}}
\def\Pc{{\cal P}}

\def\ra{\Rightarrow}

\def\<{\langle}
\def\>{\rangle}

\def\1{{\bf 1}}

\def\ad{{\rm Ad}}

\def\sp{{\rm sp}}

\def\tr{{\rm tr}}

\def\id{\pi}
\def\Ind{{\rm ind}}

\def\sec{{\rm sec}}
\def\aut{{\rm aut}}
\def\shom{{\Delta}}

\def\sgm{\sigma }

\def\ol{\overline}

\newtheorem{Def}{Definition}[section]{\bf}{\it}
\newtheorem{Lem}{Lemma}[section]{\bf}{\it}
\newtheorem{Cor}{Corollary}[section]{\bf}{\it}
\newtheorem{Tho}{Theorem}[section]{\bf}{\it}
\newtheorem{Pro}{Proposition}[section]{\bf}{\it}
\newtheorem{Ass}{Assumption}{\it}{\rm}
\def\btho{\begin{Tho}}
\def\bpro{\begin{Pro}}
\def\blem{\begin{Lem}}
\def\bdef{\begin{Def}}
\def\bcor{\begin{Cor}}
\def\bass{\begin{Ass}}
\def\bprf{\vskip 0.2cm\noindent {\it Proof.} }
\def\etho{\end{Tho}}
\def\epro{\end{Pro}}
\def\elem{\end{Lem}}
\def\eef{\end{Def}}
\def\ecor{\end{Cor}}
\def\eass{\end{Ass}}

\def\eprf{ q.e.d. \vskip 0.5cm\noindent} 
\begin{document}
\title{\bf On the Algebraic Theory of Soliton and Antisoliton Sectors}
\author{{\it Dirk Schlingemann} \\ 
II.Institut f\"ur Theoretische Physik, Universit\"at Hamburg} 
\maketitle
%%%%%%%%%%%%%%%%%%%%%%%%%%%%%%%%%%%%%%%%%%%%%%%%%%%%%%%%%%%%%%%%%%%%%%%%%%%%
\abstract{
We consider the properties of 
massive one particle states on a 
translation covariant Haag-Kastler net in Minkowski space. 
In two dimensional theories,
these states can be interpreted as 
soliton states and we are interested in the existence of 
antisolitons. It is shown that 
for each soliton state there are three different possibilities 
for the construction of an antisoliton sector which are equivalent 
if the (statistical) dimension of the corresponding 
soliton sector is finite. }
%\vskip 0.5cm\noindent
%%%%%%%%%%%%%%%%%%%%%%%%%%%%%%%%%%%%%%%%%%%%%%%%%%%%%%%%%%%%%%%%%%%%%%%%%%%%
%   19/04/94 408151423  MEMBER NAME  GTPUBA   (TEXT)     M  TEX
%\vskip 0.5cm\noindent
%%%%%%%%%%%%%%%%%%%%%%%%%%%%%%%%%%%%%%%%%%%%%%%%%%%%%%%%%%%%%%%%%%%%%%%%%%%%
\section*{Introduction}
The basic philosophy of algebraic quantum field theory 
is that the whole information about a physical system
can be obtained from the observables, representing physical
operations, and a special class of states, representing
possible preparations of the physical system.

The physical operations are described by a
Haag-Kastler net of C*-algebras in $d$-dimensional
space-time, $\Oc\mapsto\Ag(\Oc)$ \cite{DHRI,DHRII,Ha}. We denote
the smallest C*-algebra containing all
algebras $\Ag(\Oc)$ by $\Ag$. 
The elements in $\Ag(\Oc)$ are called local observables and
the elements in $\Ag$ are called quasi-local observables.
Furthermore, we select a class of states to
consider aspects which are relevant in elementary particle physics. 
A state that describes a massive particle alone in the world 
(massive one-particle state) $\omega$ is characterized by
its spectral properties, i.e. the translations
$x\mapsto\alpha_x$ are implemented in the GNS-representation of
$\omega$ by unitary operators $U_\omega(x)$, and 
the spectrum of $U_\omega$
consists of the 
mass shell $H_m=\{p:p^2=m^2\}$ and a subset of the continuum $C_{m+\mu}=
\{p:p^2>(m+\mu)^2\}$. For a massive vacuum state, we have
a different spectrum condition. Here, the spectrum consists of 
the value $0$ and a subset of the continuum $C_\mu$ \cite{BuFre}.
The positive number $\mu>0$ is called the mass gap.

It is shown by D.\ Buchholz and K.\ Fredenhagen that
for quantum field theories in $d>1+1$ space-time dimensions
there is for any massive one particle states $\omega$ a unique related
massive vacuum state $\omega_0$. Moreover, one can find an
endomorphism $\rho$ acting on a suitable extension $\Bg$ of the
algebra of quasilocal observables $\Ag$, such that
$\rho|_\Ag$ is unitarily equivalent to the GNS-representation
of $\omega$. These endomorphisms, let us call them
BF-endomorphisms, are localized in space-like
cones $S$ in the sense that $\rho(A)=A$ for any $A\in\Ag(\Oc)$
for which $\Oc$ is contained in the space-like complement of $S$
\cite{BuFre,Ha}. 

We have to mention that
there is an older description of particle states, due to 
S.Doplicher, R.Haag and J.E.Roberts \cite{DHRI,DHRII}.
They select translation covariant
representations $\pi$ whose restrictions to an algebra that
belongs to the space-like complement of a double cone $\Oc$, 
are equivalent to the vacuum representation $\pi_0$, i.e.:
$\pi(A)=v^*\pi_0(A)v$, for each $ A\in\Ag(\Oc')$.
Here $v$ is a unitary operator which depends on the double cone $\Oc$. 
This leads to 
endomorphisms $\rho:\Ag\to\Ag$, let us call them DHR-endomorphisms,
which are localized in 
double cones.   

In two dimensional
quantum field theories the massive one particle
states can have soliton properties \cite{Fre1,Fre2,Schl1},
i.e. for each massive one particle state
$\omega$ there are {\em two} massive vacuum states, 
a left vacuum state $\omega_-$ and a right vacuum state $\omega_+$, such that
$$
\lim_{|x|\to\pm\infty}\omega(\alpha_xA)=\omega_\pm(A)\ \ \ .
$$
Here we have set for $x\in\Rl^2$, $|x|:=x^1-|x^0|$ and for 
$|x|\to\pm\infty$ converges $x$ to $\pm$-space-like infinity.
The sectors belonging to the
states $\omega_\pm$ are called the asymptotic vacua of $\omega$.

It turns out that the most natural description of
superselection sectors in two dimensional field theories
is formulated in terms of algebra homomorphisms.
For any vacuum sector $a$ of $\Ag$ one obtains, in a natural way
extensions $\Ag^\pm_a$ of the $C^*$-algebra $\Ag$ depending on
the space-like directions $|x|\to\pm\infty$.
For each massive one particle state $\omega$ with left vacuum $a$ and 
right vacuum $b$, one obtains *-homomorphisms
$\rho_+:\Ag^+_b\to\Ag^+_a$ and $\rho_-:\Ag^-_a\to\Ag^-_b$, such that
$\rho_\pm|_\Ag$ is unitarily equivalent to the GNS-representation
of $\omega$. These *-homomorphisms, let us call them
one-soliton homomorphisms, are localized in space-like wedges
of the type
$W_\pm+x$, where $W_\pm$ are the regions given by
$W_\pm:=\{x\in\Rl^2:|x^0|<\pm x^1\}$.
Thus there are two types of soliton homomorphisms, namely 
homomorphisms with orientation $q=-$, i.e. 
$\rho_-$ is localized in a region $W_-+x$,
and homomorphisms with orientation $q=+$, i.e.
$\rho_+$ is localized in a region $W_++x$. 
Hence for each soliton superselection sector 
%%%%%%%%%%%%%%%%%%%%%%%%%%%%%%%%%%%%%%%%%%%%%%%%%%%%%%%%%%%%%%%%%%%%%%%%%%%%%%%
\footnote{A soliton superselection sector is a class
$\theta=[\rho]$ of unitarily equivalent soliton homomorphisms
$\rho_1\cong\rho_2$, i.e. the representations $\rho_1|_\Ag$ and 
$\rho_2|_\Ag$ are unitarily equivalent.}
%%%%%%%%%%%%%%%%%%%%%%%%%%%%%%%%%%%%%%%%%%%%%%%%%%%%%%%%%%%%%%%%%%%%%%%%%%%%%%
one can find a representative $\rho_+$ with orientation $q=+$ and 
a representative $\rho_-$ with orientation $q=-$.  

In contrast to the DHR- and BF-case, it is not clear whether two soliton 
sectors can be always composed, but it is possible to compose 
soliton sectors $\theta_1$ and $\theta_2$, if the left vacuum 
corresponding to $\theta_1$ is equal to the right vacuum of
$\theta_2$.
To see this, we choose soliton homomorphisms $\rho_1:\Ag_a^+\to\Ag_b^+$, 
$\rho_1':\Ag_b^-\to\Ag_a^-$, representing the sector $\theta_1$ and 
soliton homomorphisms $\rho_2:\Ag_c^+\to\Ag_d^+$, $\rho_2':\Ag_d^-\to\Ag_c^-$,
representing the sector $\theta_2$. 
The left vacuum of $\theta_1$ is $b$ and the right vacuum of 
$\theta_2$ is $c$. For $c=b$ we may compose $\rho_2$ and
$\rho_1$
as well as $\rho_1'$ and $\rho_2'$.
Thus there are two possibilities for the composition of soliton sectors, namely
$$
\theta_1\times\theta = [\rho_1]\times [\rho] := [\rho_1\rho] \ \ \ {\rm and}
\ \ \ 
\theta\diamond\theta_1 = [\rho']\diamond [\rho_1'] := [\rho'\rho_1']
$$
It is shown in \cite{Fre2} that 
the homomorphisms $\rho_1\rho$ and $\rho'\rho_1'$
represent the same superselection sector.
Hence both ways to 
compose soliton sectors are equivalent, i.e.: 
$$
\theta_1\times\theta = \theta\diamond\theta_1
$$

The structure of soliton superselection sectors is discussed 
in section {\bf 1}.

We will see in section {\bf 2} that there are different ways to obtain 
an antisoliton sector.
First, one can use the methods developed by 
D. Guido and R. Longo \cite{GuiLo}. 
Let $J_a$ be the modular conjugation with respect to the 
von Neumann algebra $\pi_a(\Ag(W_+))''$ and the cyclic and 
separating vector $\Omega_a$. 
%%%%%%%%%%%%%%%%%%%%%%%%%%%%%%%%%%%%%%%%%%%%%%%%%%%%%%%%%%%%%%%%%%%%%%%%%%%%%%%
\footnote{ For each vacuum sector $a$ we choose a 
corresponding vacuum state $\omega_a$. 
$(\Hc_a,\pi_a,\Omega_a)$ 
is its GNS-representation. 
We consider the weak closure of $\pi_a(\Ag(W_\pm+x))$ in 
the algebra of bounded operators $\Bc(\Hc_a)$ that is denoted by
$\pi_a(\Ag(W_\pm+x))''$. The vector $\Omega_a$ is cyclic for 
all algebras $\pi_a(\Ag(W_\pm+x))''$ and hence cyclic and separating
for $\pi_a(\Ag(W_+))''$. Thus one has modular data 
$\Delta_a,J_a$ with respect to the pair $(\pi_a(\Ag(W_+))'',\Omega_a)$.}
%%%%%%%%%%%%%%%%%%%%%%%%%%%%%%%%%%%%%%%%%%%%%%%%%%%%%%%%%%%%%%%%%%%%%%%%%%%%%%
We define $j_a:=\ad(J_a)$ and interpret
$j_a\circ\rho\circ j_b:\Ag_a^-\to\Ag_b^-$ as an antisoliton homomorphism.
To show that $j_a\circ\rho\circ j_b$ is localized in a wedge region,
we assume that the net $\Oc\mapsto\Ac(\Oc)$ is equipped 
with a PCT-symmetry, i.e.  
there is an antiautomorphism $j:\Ag\to\Ag$, 
that is implemented in each vacuum sector $a$ by the modular conjugation 
$J_a$.
If $\rho$ is localized in the wedge region $W_++x$, one finds that
$j_a\circ\rho\circ j_b$ is localized in the PT-reflected 
region $W_--x$.
 
For our purpose, it is important to consider 
compositions of soliton and antisoliton sectors and we wish to 
obtain an antisoliton homomorphism for $\rho$ that is localized in the 
same region as $\rho$. For this reason, we choose a 
soliton homomorphism $\rho':\Ag_b^-\to\Ag^-_a$ that 
is localized in $W_--x$ and represents the same sector as $\rho$.
We shall show that $j_a\circ\rho\circ j_b$ and $j_b\circ\rho'\circ j_a$
belong to the same superselection sector and we can also choose 
$\bar\rho_0:=j_b\circ\rho'\circ j_a:\Ag^+_b\to\Ag_a^+$ as an antisoliton 
homomorphism.
We call $\bar\rho_0$ a $C_0$-conjugate for $\rho$ and shall
prove that $\bar\rho_0$ has the following properties:

For a soliton homomorphism $\sgm$ which is unitarily equivalent to
$\rho$, 
the homomorphisms $\bar\sgm_0$ and $\bar\rho_0$ are unitarily equivalent.
The state $\omega_b\circ\bar\rho_0$ 
is a massive one particle state possessing  
the same mass spectrum as $\rho$.

To motivate another method for the construction of antisoliton sectors,
let us return to the $d>1+1$ dimensional case for which K.\ Fredenhagen 
\cite{Fre3} has constructed antiparticle sectors. 
In order to understand the physical idea behind his method, 
let us consider a particle-antiparticle state which 
is a state in the vacuum sector. Now suppose that   
$\omega$ is of the form
$$
\omega(A)=\<\psi,\pi_0(A)\psi\> \ \ \  ,
$$ 
where $\psi$ is a vector in the Hilbert space $\Hc_0$ of the vacuum 
representation $\pi_0$.
Since the vacuum can be approximated by shifting a one particle state to 
space-like infinity \cite{BuFre},
one obtains
$$
\omega(A)=\lim_{|x|\to\infty}\<\psi,\rho(\alpha_xA)\psi\>
=\lim_{|x|\to\infty}\<\psi,U_\rho(x)\rho(A)U_\rho(-x)\psi\> \ \ \ .
$$  
Now for observables $A$ that are localized in the 
space-like complement of the localization region of $\rho$ one gets
$$
\omega(A)=\lim_{|x|\to\infty}\<\psi,U_\rho(x)AU_\rho(-x)\psi\> \ \ \ .
$$  
The particle which is described by $\rho$ is outside this region
and what is left is an antiparticle state.
It is shown in \cite{Fre3} that this state can also be described by a
endomomorphism $\bar\rho$, i.e. there is 
a vector $\psi'\in\Hc_a$ such that:
$$
\<\psi',\bar\rho(A)\psi'\>
=\lim_{|x|\to\infty}\<\psi,U_\rho(x)AU_\rho(-x)\psi\> 
$$  
The endomorphism $\bar\rho$ can be chosen to be 
localized in the same region as $\rho$. 
Furthermore, by construction $\bar\rho\rho$ contains the 
vacuum representation. Using the statistics operator for 
$\rho$ and $\bar\rho$ we can conclude that 
$\rho\bar\rho$ too contains the vacuum representation and
the statistical dimension of $\rho$ is finite.
Before we return to the two dimensional situation, 
let us briefly summarize the results of \cite{Fre3}:

Let $\Oc\mapsto\Ag(\Oc)$ be a quantum field theory in
$d>1+1$ dimensions. Then there exists for each massive
one-particle endomorphisms $\rho$ a massive
one-particle endomorphism $\bar\rho$, such that
(1) $\sp(U_\rho)=\sp(U_{\bar\rho})$ and (2) both
$\rho\bar\rho$ and $\bar\rho\rho$ contain the
vacuum representation $\pi_0=\id$ precisely once.

Indeed, one can interpret $\bar\rho$ as an antiparticle for
$\rho$. Property (1) shows that
the antiparticle has the same mass as the particle and
property (2) shows that the antiparticle
carries the inverse charge of the particle with
respect to the vacuum $\pi_0$.

The methods used by K. Fredenhagen can be also applied to  
the two dimensional situation, but here one has two possibilities 
to shift the isolated particle to space-like infinity. 
Firstly, for a soliton homomorphism $\rho:\Ag_a^+\to\Ag^+_b$,
one can consider the limit at positive space-like infinity
$$
\<\psi_+,\bar\rho_+(A)\psi_+\>
=\lim_{|x|\to +\infty}\<\psi,U_\rho(x)AU_\rho(-x)\psi\> 
$$
Second, one can also consider the limit at negative space-like infinity
$$
\<\psi_-,\bar\rho_-(A)\psi_-\>
=\lim_{|x|\to -\infty}\<\psi,U_\rho(x)AU_\rho(-x)\psi\> \ \ \ .
$$
We shall show that in the two dimensional case 
one obtains two soliton homomorphisms $\bar\rho_+:\Ag_b^+\to\Ag_a^+$
and $\bar\rho_-:\Ag_b^+\to\Ag_a^+$, let us call 
them $C_\pm$-conjugates for $\rho$, with 
the following 
properties:

The homomorphisms $\bar\rho_+$ and $\bar\rho_-$ are  
translation covariant and can be localized in a wedge 
region $W_++x$. Moreover,
the homomorphism $\bar\rho_+\rho$ contains the 
vacuum representation $\pi_a$ and the 
homomorphism $\rho\bar\rho_-$ contains the vacuum representation $\pi_b$.
%By this properties, $\bar\rho_+$ and $\bar\rho_-$ 
%are fixed uniquely up to unitary equivalence.
 
In contrast to the DHR- and BF-case, we can not conclude that 
$\rho\bar\rho_+$ contains the vacuum representation
$\pi_b$ and that $\bar\rho_-\rho$ contains the vacuum representation 
$\pi_a$, because for 
the soliton case there is no analogue of a statistics operator. 
Moreover, it is not clear whether 
$\bar\rho_+$ and $\bar\rho_-$ are equivalent or not. 

In section {\bf 3} we discuss the relations between the different 
constructions.
We shall show that the $C_0$-conjugate of a $C_+$-conjugate
for a soliton homomorphism $\rho$ 
is equivalent to the $C_-$-conjugate of a $C_0$-conjugate 
for $\rho$, i.e.:
$$
j_a\circ(\bar\rho_+)\circ j_b \cong \ol{(j_b\circ\rho\circ j_a)}_-
$$

It is shown by R. Longo that for a 
DHR- or BF-endomorphism $\rho$ the index of the
inclusion $\rho(\Ag(S))\subset\Ag(S)$, here $S$ denotes 
region where $\rho$ is localized, is precisely the square
of the statistical dimension of $\rho$ \cite{Lo1,Lo2}.
Analogously, the square of the
dimension of a soliton homomorphism $\rho:\Ag^\pm_a\to\Ag^\pm_b$ 
is {\em defined} by the
index of the inclusion
$\rho(\Ag(W_\pm+x)_a)\subset\Ag(W_\pm+x)_b$, i.e.:
\be
d(\rho)^2=\Ind(\rho(\Ag(W_\pm+x)_a),\Ag(W_\pm+x)_b)
\ee
Here $W_\pm+x$ is the localization
region of $\rho$.

We shall prove  
that the dimension of a soliton homomorphism 
$\rho:\Ag_a^\pm\to\Ag_b^\pm$ is finite if and
only if there exists a homomorphism
$\bar\rho:\Ag_b^\pm\to\Ag_a^\pm$ such that
$\bar\rho\rho$ contains the 
vacuum representation $\pi_a$ and
$\rho\bar\rho$ contains the vacuum representation $\pi_b$.
We mention that 
this is a slight generalization of R. Longo's result in \cite{Lo2}
and we are using the same methods to prove it.

Suppose a massive one
soliton homomorphism $\rho:\Ag_a^\pm\to\Ag_b^\pm$ fulfills one  
of the following three
conditions:
\begin{description}
\item[{\it 1:}]
The dimension of $\rho$ is finite.
\item[{\it 2:}]
The $C_+$-conjugate and the $C_-$-conjugate of
$\rho$ are equivalent.
\item[{\it 3:}]
Either, the $C_+$- or the $C_-$-conjugate and the $C_0$-conjugate
are equivalent.
\end{description}
Then we will show that there exists a soliton homomorphism 
$\bar\rho:\Ag_b^\pm\to\Ag_a^\pm$ with the following properties:

The homomorphism $\bar\rho$ is 
by one of the properties 1-3, up to unitary equivalence,
uniquely determined.
The state $\omega_a\circ\bar\rho$ is a massive one particle state
possessing the same mass spectrum as $\rho$.
Furthermore, $\bar\rho\rho$ contains the 
vacuum representation $\pi_a$ and
$\rho\bar\rho$ contains the vacuum representation $\pi_b$.  

%%%%%%%%%%%%%%%%%%%%%%%%%%%%%%%%%%%%%%%%%%%%%%%%%%%%%%%%%%%%%%%%%%%%%%%%%%%%%%
\section{Soliton Sectors}
%%%%%%%%%%%%%%%%%%%%%%%%%%%%%%%%%%%%%%%%%%%%%%%%%%%%%%%%%%%%%%%%%%%%%%%%%%%%%%
In this section, we discuss the mathematical structure of
soliton sectors from the algebraic point of view.
Based on the work of
K. Fredenhagen \cite{Fre2}, we describe 
the construction of soliton homomorphisms which are the
analogue of the charged endomorphisms in the DHR- and BF-analysis
\cite{DHRII,BuFre}.

Before we start our discussion, we give some mathematical preliminaries.
Let us denote by 
$W_\pm\subset\Rl^2$ the {\sl wedge region} $\{(t,x):|t|<\pm x\}$. 
A {\sl double cone} $\Oc\subset\Rl^2$ 
is a non-empty intersection of a {\sl left wedge region } $W_-+x$
and a {\sl right wedge region} $W_++y$, $\Oc=W_-+x\cap W_++y$.
The collection of all double cones is denoted by $\Kc$.

We list now the axioms which should be fulfilled by the theories 
under consideration.    
\begin{description}
\item[Axiom I:]
There is a net of $C^*$-algebras, i.e. a prescription            
that assigns to each double cone $\Oc\in\Kc$ a $C^*$-algebra $\Ag(\Oc)$,
called {\sl algebra of local observables}. 
Furthermore, this prescription is 
{\sl isotonous}, i.e.
$$
\Oc\subset\Oc_1 \ \ \ \ra \ \ \ \Ag(\Oc)\subset\Ag(\Oc_1) \ \ \, 
$$
and {\sl local}, i.e.
$$
\Oc\subset\Oc_1' \ \ \ \ra \ \ \ [\Ag(\Oc),\Ag(\Oc_1)]=\{0\} \ \ \,
$$
where $\Oc_1'$ is the space-like complement of $\Oc_1$.
We call the $C^*$- algebra
$$
\Ag:=\ol{\bigcup_{\Oc\in\Kc}\Ag(\Oc)}^{||\cdot ||} \ \ \ .
$$
the {\sl algebra of quasi local observables}. Moreover, suppose 
there exists a group homomorphism $\alpha$ from
the two dimensional translation group $T(\Rl^2)\cong\Rl^2$ into the 
automorphism group $\aut(\Ag)$ of $\Ag$, such that
$$
\alpha_x(\Ag(\Oc))\subset\Ag(\Oc+x) \ \ \ ,
$$
for each $x\in\Rl^2$.
Such a prescription $\Oc\in\Kc\mapsto\Ag(\Oc)$ is called a
two-dimensional {\sl Haag-Kastler-net} with translation
covariance.
\item[Axiom II:]
There are {\sl translation covariant states} on 
$\Ag$. These are states $\omega$ such that there is a unitary
strongly continuous representation $U_\omega$ of the translation 
group on the GNS-Hilbert space $\Hc_\omega$ of $\omega$, with the 
property
$$
\pi_\omega(\alpha_xA)=U_\omega(x)\pi_\omega(A)U_\omega(x)^* \ \ \ ;
\ \ \ \forall A\in \Ag ; \ \ \ \forall x\in\Rl^2 \ \ \  .
$$
Here $(\Hc_\omega,\pi_\omega,\Omega_\omega)$ is the GNS-triple,
belonging to $\omega$.
Moreover, the set of {\sl massive one-particle} and {\sl massive
vacuum states} is not empty. A massive one-particle state 
is a {\sl pure translation covariant state} on $\Ag$, 
such that the spectrum $\sp(U_\omega)$ of the generators of 
the translation group $U_\omega$ is consists of
the {\sl mass shell} $H_m=\{p:p^2=m^2\}$ and a subset of
the continuum $C_{m+\mu}=\{p:p^2>(m+\mu)^2\}$.
For a massive vacuum state, we have
a different {\sl spectrum condition}. Here, the spectrum consists of
the value $0$ and a subset of $C_\mu$ \cite{BuFre}.
The positive number $\mu>0$ is called the {\sl mass gap}. 
\end{description}

We say that two states $\omega,\omega_1$ belong to the same {\sl sector} if
their GNS-representations $\pi_\omega,\pi_{\omega_1}$ are 
unitarily equivalent, i.e. there is a unitary operator 
$u:\Hc_\omega\to\Hc_{\omega_1}$,
such that $u\pi_\omega(A)=\pi_{\omega_1}(A)u$, for each $A\in\Ag$.
A state $\omega_1$ belongs to the {\sl folium} of a state $\omega$ if
$\omega_1$ is of the form $\omega_1(A)=\tr(T\pi_\omega(A))$ for each $A\in\Ag$,
where $T$ is a trace class operator in the algebra $\Bc(\Hc_\omega)$
of bounded operators on $\Hc_\omega$.

Let us denote the collection of all {\sl massive
vacuum sectors} by $\sec_0$. For each sector $a\in\sec_0$, we
fix one massive vacuum state $\omega_a$ and write
$(\Hc_a,\pi_a,\Omega_a)$ for the GNS-triple of $\omega_a$.
%We define now the wedge $C^*$-algebras 
%$$
%\Ag(W_\pm+x):=
%\ol{\bigcup_{\Oc\subset W_\pm+x}\Ag(\Oc)}^{||\cdot ||} \ \ \ .
%$$
%For later purpose, 
%we define for each vacuum sector $a\in\sec_0$ the
%following von-Neumann-algebras:
%$$
%\Ag(W_\pm+x)_a:=\pi_a(\Ag(W_\pm+x))''
%$$

For later purpose, let us define the following algebras:
\begin{description}
\item[{\rm (a)}]
The wedge C*-algebras  
$$
\Ag(W_\pm+x):=
\ol{\bigcup_{\Oc\subset W_\pm+x}\Ag(\Oc)}^{||\cdot ||} \ \ \ .
$$
\item[{\rm (b)}]
The von-Neumann-algebras with respect to the vacuum $a\in\sec_0$
$$
\Ag(W_\pm+x)_a:=\pi_a(\Ag(W_\pm+x))'' \ \ \ .
$$
\end{description}
 
Without loss of generality, we may assume that the representation 
$\pi_a$ is faithful \cite{Fre2}. Thus we can
interpret $\Ag(W_\pm+x)$ as a common weakly dense subalgebra of
the von-Neumann-algebras $\Ag(W_\pm+x)_a$.
Obviously, the algebras
$$
\Ag^\pm_a:=\ol{\bigcup_{x\in\Rl^2}\Ag(W_\pm+x)_a}^{||\cdot ||}
$$
are extensions of the C*-algebra of quasi-local observables $\Ag$ and
the vacuum representation $\pi_a$ has  canonical extensions
$\pi_a^\pm$ to the $C^*$-algebras $\Ag^\pm_a$, for which we
write shortly $\id_a$ instead of $\pi_a^\pm$.
\bdef
A state $\omega$ on $\Ag$ is called {\sl  a soliton state}, if it
satisfies the following conditions:
\begin{description}
\item[{\it 1:}]
The state $\omega$ is translation covariant.
\item[{\it 2:}]
There are vacuum sectors $a,b\in\sec_0$ and
unitary operators $v:\Hc_\omega\to\Hc_b$ and
 $v':\Hc_\omega\to\Hc_a$, such that:
\begin{equation}
\pi_\omega(A')={v'}^*\id_a(A')v' \ \ \ {\rm and} \ \ \
\pi_\omega(A)=v^*\id_b(A)v
\end{equation}
for all $A\in\Ag(W_+)$ and for all $A'\in\Ag(W_-)$.
\end{description}
The vacuum sectors $a,b$ are called asymptotic vacua associated with 
the state $\omega$.
\eef
One can deduce from the properties of soliton states
the following statement \cite{Fre2}: 
\blem
Let $\omega$ be a soliton state, $a,b\in\sec_0$
the asymptotic vacua as in Definition 1.1 above and
$\pi_\omega$ its GNS-representation.
Then there are unique
extensions of $\pi_\omega$,
$$
\pi_\omega^+:\Ag_b^+\to\Bc(\Hc_\omega) \ \ \ {\rm and } \ \ \ 
\pi_\omega^-:\Ag_a^-\to\Bc(\Hc_\omega) \ \ \ .
$$
\elem

For a soliton state $\omega$ with asymptotic vacua $a,b$
as above,
we define *-homomorphisms
$$
\rho=\ad(v')\circ\pi_\omega^+:
\Ag_b^+\to\Bc(\Hc_a)
\ \ \ {\rm and } \ \ \ 
\rho'=\ad(v)\circ\pi_\omega^-:
\Ag_a^-\to\Bc(\Hc_b)
$$
where $v,v'$ are given as in (1). By construction, 
the restrictions $\rho|_\Ag$ and $\rho'|_\Ag$
are unitarily equivalent. To obtain 
homomorphisms that map $\Ag^+_b$ into $\Ag^+_a$
or $\Ag^-_a$ into $\Ag^-_b$ we have to make an assumption. 
\bass
The algebras $\Ag(W_\pm+x)_a$ fulfill
duality for wedges for all vacuum sectors $a\in\sec_0$, namely
$$
\Ag(W_\pm+x)_a=\Ag(W_\mp+x)_a' \ \ \ .
$$
\eass

It can then be shown by combining Assumption 1 with the methods in
\cite{Fre2} that the image of $\rho$ is contained
in $\Ag_a^+$ and that the image of $\rho'$ is contained
in $\Ag_b^-$.

By construction, the *-homomorphism $\rho$ is localized in
$W_++x$, i.e.:
$$
\rho|_{\Ag(W_-+x)}=\id_{\Ag(W_-+x)}
$$
Moreover, $\rho$ is implemented by a unitary operator 
$u:\Hc_b\to\Hc_a$ on any von-Neumann-algebra $\Ag(W_++y)_b$.
Using equation (1), one sees that for each $A\in\Ag(W_++x)_b$
one has
$$
\rho(A)=v'v^*Av{v'}^*  \ \ \ .
$$
Since $\rho$ is translation covariant, it is implemented in such a way
on each algebra $\Ag(W_++y)_b$.
Of course, the same is true if one exchanges $\rho$ by $\rho'$,
$a$ by $b$ and the $+$ by the $-$ sign.

From the discussion above, we see that
for each soliton state $\omega$ there are two types 
of {\sl soliton homomorphisms},
which represent the same superselection sector. The homomorphism $\rho$ 
is localized in a right wedge region $W_++x$, describing the creation 
of a soliton charge $[\rho]$
out of the vacuum $a$ and connecting it to
the vacuum $b$. On the other hand, 
$\rho'$ is localized in a left wedge region, 
describing the creation of the same charge $[\rho]$
out of the vacuum $b$ and connecting it to
the vacuum $a$.
In the following, we use the notation
$\rho:a\to b$, $\rho':b\to a$.

Now, consider a soliton state $\omega$, with asymptotic vacua
$a,b$ and a soliton state $\omega_1$ with asymptotic vacua $c,d$.
For each of the two states $\omega,\omega_1$, we obtain a 
*-homomorphism localized in a left wedge and a *-homomorphism 
localized in a right wedge, i.e.: 
$$
\begin{array}{l}
{\rm For} \ \ \omega \ \ {\rm  \ \ we  \ \ obtain \ \ } 
\rho:\Ag_b^+\to\Ag^+_a \ \ \ {\rm and} \ \ \
\rho':\Ag_a^-\to\Ag^-_b \ \ .
\vspace{0.2cm}\\
{\rm For} \ \ \omega_1 \ \ {\rm  \ \ we  \ \ obtain \ \ } 
 \rho_1:\Ag_d^+\to\Ag^+_c \ \ \ {\rm and} \ \ \
\rho'_1:\Ag_c^-\to\Ag^-_d \ \ .
\end{array}
$$
For $b=c$, the algebra $\rho_1(\Ag_d^+)$
is contained in $\Ag_b^+$ and 
$\rho'(\Ag_a^-)$ is contained in $\Ag_c^-$. Thus we may  compose
$\rho$ with $\rho_1$ and $\rho_1'$ with $\rho'$ and obtain further 
*-homomorphisms
$$
\rho\rho_1:\Ag_d^+\to\Ag^+_a \ \ \  {\rm and} \ \ \
\rho_1'\rho':\Ag_a^-\to\Ag^-_d \ \ \ .
$$
The proof of the following statement can be found in \cite{Fre2}. 
\bpro
Both representations
$$
\rho\rho_1|_\Ag  \ \ \ {\rm and} \ \ \
\rho_1'\rho'|_\Ag
$$ 
are unitarily equivalent.
\epro

%%%%%%%%%%%%%%%%%%%%%%%%%%%%%%%%%%%%%%%%%%%%%%%%%%%%%%%%%%%%%%%%%%%%%%%%%%%%%%%%
From
the results of \cite{BuFre,Fre2,Schl1} we may conclude that there
is a special class of soliton states, namely states with
particle character. This fact can be formulated by the following 
proposition:
%%%%%%%%%%%%%%%%%%%%%%%%%%%%%%%%%%%%%%%%%%%%%%%%%%%%%%%%%%%%%%%%%%%%%%%%%%%%%%%%%%
\bpro
If $\omega$ is a massive one-particle state, then $\omega$ is a soliton state.
\epro
%%%%%%%%%%%%%%%%%%%%%%%%%%%%%%%%%%%%%%%%%%%%%%%%%%%%%%%%%%%%%%%%%%%%%%%%%%%%%%%%

We call massive one-particle states simply {\sl one soliton states}
and the corresponding homomorphisms {\sl one-soliton homomorphisms}.

For $q=\pm$, let us denote by $\shom(q,x)$, the set of all
soliton homomorphisms which are localized in $W_q+x$. In addition to that,
the set of one-soliton homomorphisms is denoted by $\shom_1(q,x)$.

For each soliton homomorphism $\rho:a\to b$ 
contained in $\shom(q,x)$,
we define the {\sl source} $s(\rho):=a$ and the
{\sl range} $r(\rho):=b$.
The value $q$ is called the {\sl orientation} of a soliton
homomorphism. 

A soliton homomorphism $\rho_1$ is called a {\sl subobject} of a 
soliton homomorphism $\rho$, if there exists an 
isometric intertwiner $v$ from $\rho_1$ to $\rho$, i.e.:
$$
v\rho_1(A)=\rho(A)v \ \ \ ; \ \ \ A\in\Ag
$$

Furthermore, one has $s(\rho)=s(\rho_1)$ and
$r(\rho)=r(\rho_1)$ if $\rho_1$ is a subobject of $\rho$.
Conversely, we are able to define the {\sl direct sum} 
$\rho_1\oplus\rho_2$, for each pair of soliton homomorphisms 
$\rho_1,\rho_2$ with $s(\rho_1)=s(\rho_2)$ and $r(\rho_1)=r(\rho_2)$.
According to \cite{Bor3} we can find two  
isometries $v_1,v_2\in \Ag(W_q+x)_a$ with complementary range
and define $\rho_1\oplus\rho_2$
as follows:
$$
\rho(A)=
(\rho_1 \oplus \rho_2 )(A) := v_1\rho_1(A)v_1^*+v_2\rho_2(A)v_2^*
$$ 
If $W_q+x$ contains the localization regions of $\rho_1$ and $\rho_2$,
$\rho$ is a soliton homomorphism which is also localized in $W_q+x$.
\blem

\begin{description}
\item[{\it 1:}]
If $\rho$ is a soliton homomorphism and 
$\pi_1$ a subrepresentation of $\rho|_\Ag$, i.e.
there is an isometry $v:\Hc_{\pi_1}\to\Hc_{s(\rho)}$ with
$$
v\pi_1(A)=\rho(A)v \ \ \ ; \ \ \ A\in\Ag \ \ \ ,
$$
then there exists a subobject $\rho_1$ of $\rho$, such that 
$\rho_1|_\Ag$ is unitarily equivalent to $\pi_1$.
\item[{\it 2:}]  
For each pair of soliton homomorphisms
$\rho_1,\rho_2\in\shom(q,x)$, the space of intertwiners
$I(\rho_1,\rho_2)$ contains nontrivial elements, only if
the sources and ranges of $\rho_1$ and $\rho_2$ are equal.
$$
I(\rho_1,\rho_2)\not=\{0\} \ \ \ra \ \
s(\rho_1)=s(\rho_2) \ \ {\rm and} \ \ r(\rho_1)=r(\rho_2)
$$
\end{description}
\elem
\bprf
{\it 1}: We choose $\rho:a\to b$ to be a soliton homomorphism, 
localized in $W_++x$ and
$\rho':b\to a$ a soliton homomorphism localized in $W_-+x$, such that 
the representations $\rho|_\Ag$ and $\rho'|_\Ag$ are unitarily
equivalent. As mentioned above, such a choice is always possible.
For a subrepresentation $\pi_1$ of $\rho|_\Ag$ there are 
isometries $v:\Hc_{\pi_1}\to \Hc_a$ and 
$v':\Hc_{\pi_1}\to \Hc_b$ such that
$$
v\pi_1(A)=\rho(A)v \ \ \ {\rm and}  \ \ \ v'\pi_1(A)=\rho'(A)v' \ \ \ .
$$
The projection $E=vv^*$ is contained in $\Ag(W_++x)_a$
and the projection $E'=v'{v'}^*$ is contained in $\Ag(W_-+x)_b$,
since $\rho$ is localized in $W_++x$ and $\rho'$ is localized in
$W_-+x$.
According to \cite{Bor3} there are isometries 
$w\in\Ag(W_++x)_a$ and $w'\in\Ag(W_-+x)_b$,
such that $E=ww^*$ and $E'=w'{w'}^*$.
Now we define the unitaries $v_1':=w^*v$ and $v_1:={w'}^*v'$
and obtain for $A\in\Ag(W_++x)$ and $A'\in\Ag(W_-+x)$:
$$
\pi_1(A)=v_1^*\pi_b(A)v_1 \ \ \ {\rm and} \ \ \ \pi_1(A')={v_1'}^*\pi_a(A')v_1'
$$
Thus $\pi_1$ is a soliton representation which implies that 
there is a subobject for $\rho$ which is unitarily equivalent to $\pi_1$.

{\it 2}:
Suppose there are soliton homomorphisms 
$\rho_1:a\to b$ and $\rho_2:c\to d$, 
such that there is an nontrivial intertwiner $v:\Hc_a\to\Hc_c$
that intertwines the representations $\rho_1|_\Ag$ and $\rho_2|_\Ag$.
If $\rho_1$ and $\rho_2$ have the same orientation, then 
we can conclude by {\it 1}
that there is a joint subobject $\rho$ for $\rho_1$ and $\rho_2$.
Now we obtain
$$
a=s(\rho)=c \ \ \ {\rm and} \ \ \ b=r(\rho)=d
$$
and the proof is complete.
\eprf
\bcor
The set of soliton homomorphisms  $\shom(q,x)$ is 
closed under multiplication, taking direct sums and subobjects.
\ecor
\bdef
Two soliton homomorphisms $\rho\in\shom(q,x)$
and $\rho\in\shom(p,y)$ are called {\sl inner
unitarily equivalent}, if $p=q$ and there exists a unitary
intertwiner $u\in I(\rho,\rho_1)$.
We call $\rho$ and $\rho_1$ {\sl unitarily equivalent}, if
the representations $\rho|_\Ag$ and $\rho_1|_\Ag$ are
unitarily equivalent.
We denote the set of inner unitary equivalence classes by
$\sec(q)$ and the set of unitary equivalence classes by
$\sec$.
We call the elements of $\sec(q)$ {\sl inner soliton sectors}
and the elements of $\sec$ {\sl soliton sectors}.
The projection which maps a soliton homomorphism onto its
inner unitary equivalence class is denoted by
$e:\rho\in\shom(q,x)\mapsto e(\rho)\in\sec(q)$,
the projection which maps a soliton homomorphism onto its
unitary equivalence class is denoted as usual by
$[\cdot]:\rho\in\shom(q,x)\mapsto [\rho]\in \sec$.
\eef
According to Lemma 1.2, we may define the source and range maps and
the composition for inner soliton sectors:
$$
\begin{array}{l}
s(e(\rho)):=s(\rho) \ \ \ \ r(e(\rho)):=r(\rho)
\vspace{0.2cm}\\
e(\rho_1)e(\rho_2):=e(\rho_1\rho_2) \ \ \ {\rm for} \ \
s(\rho_2)= r(\rho_1)
\end{array}
$$
The inner sector which belongs to the identity is
denoted by $i_a:=e(\id_a)$, $a\in\sec_0$.

Of course, the inner unitary equivalence is stronger
than the unitary equivalence, because
the inner unitary equivalence does not forget
the information about the {\sl orientation} of a
soliton homomorphism whereas the unitary equivalence
only sees the {\sl charge } of a soliton homomorphism.

Both, the set of soliton homomorphisms $\shom(q,x)$ and
the set of inner soliton sectors $\sec(q)$ are naturally
equipped with a category structure.
The objects are the vacuum sectors and the arrows are
the soliton homomorphisms or inner soliton sectors.
Moreover, there is a second natural category structure, namely 
the objects are soliton homomorphisms $\rho,\rho_1\in\shom(q,x)$
and the arrows are intertwiner $v\in I(\rho_1,\rho)$.
Such a structure is also known as a 2-$C^*$-category. 
See \cite{Rb3} for this notion.  

In the following lines we show that soliton homomorphisms
which are unitarily equivalent and have
the same orientation,
are indeed inner unitarily equivalent.
\blem
Let $\rho\in\shom(q,x)$ and $\rho'\in\shom(p,y)$ be
soliton homomorphisms with $[\rho]=[\rho']$, then for $p=q$
it follows: $e(\rho)=e(\rho')\in\sec(p=q)$.
\elem
\bprf
Since $\rho$ and $\rho'$ are localized in a $q$-wedge region,
there is $z\in\Rl^2$, such that both $\rho$ and $\rho'$ are
localized in $W_q+z$. The representations $\rho|_\Ag$ and
$\rho'|_\Ag$ are unitarily equivalent, and there is a
unitary intertwiner $u$ from $\rho$ to $\rho'$. But for
$A\in\Ag(W_{-q}+z)$ we obtain $[A,u]=0$.
Thus, $u$ is contained in $\Ag(W_q+z)_{s(\rho)}$ and we
conclude that $\rho$ and $\rho'$ are inner unitarily
equivalent.
\eprf

{\it Remark:} From the discussion above it follows that
unitarily equivalent soliton homomorphisms with
opposite orientation can not be inner equivalent.

Before we close this section, let us summarize
the mathematical structure of the set of
inner soliton sectors $\sec(q)$.

As mentioned above, one can interpret $\sec(q)$ as a
category whose objects are the vacuum sectors
$\sec_0$ and whose arrows are the inner
soliton sectors $\sec(q)$.
We write $\sec(q|a,b)$ for the set of arrows
$\theta\in\sec(q)$ with $s(\theta)=a$ and
$r(\theta)=b$.

With respect to the direct sum $\oplus$ of soliton homomorphisms, 
the sets of arrows $\sec(q|a,b)$ 
are commutative rings over the natural numbers. 

For arrows $\theta_1,\theta_2\in\sec(q|a,b)$,
$\vartheta\in\sec(q|b,c)$ and $\hat\vartheta\in\sec(q|c,a)$
the following distributive laws are fulfilled:
$$
(\theta_1\oplus\theta_2)\vartheta =\theta_1\vartheta\oplus
\theta_2\vartheta
\ \ \ {\rm and} \ \ \ 
\hat\vartheta(\theta_1\oplus\theta_2)=\hat\vartheta\theta_1\oplus
\hat\vartheta\theta_2
$$

The sets of arrows $\sec(q|a,b)$ are equipped with
a partial order relation. We write $\theta_1<\theta$ if
$\theta_1$ is a subobject for $\theta$. This relation 
satisfies $\theta_1<\theta_1\oplus\theta$.

For later purpose, it is convenient to consider
a special class of functorial operations
from $\sec(q)$ to $\sec(p)$.
\bdef
A map $j:\sec(q)\to\sec(p)$ is called a {\sl conjugation} or 
{\sl co-conjugation}
if $j$ fulfills the following conditions:
\begin{description}
\item[{\it 1 Functor, Cofunctor Condition:}]
For $s(\theta_2)=r(\theta_1)$ we have:
$$ {\rm Functor:} \ \ \ 
j(\theta_1\theta_2)=j(\theta_1)j(\theta_2)
\ \ \ {\rm or } \ \ \ {\rm Cofunctor:} \ \ \ 
j(\theta_1\theta_2)=j(\theta_2)j(\theta_1) 
$$
\item[{\it 2 Additivity:}]
For $s(\theta_1)=s(\theta_2)$ and $r(\theta_1)=r(\theta_2)$
we have:
$$
j(\theta_1\oplus\theta_2)=
j(\theta_1)\oplus j(\theta_2)
$$
\item[{\it 3 Isotony:}]
Let $\theta_1$ be a subobject of $\theta$, then
$j(\theta_1)$ is a subobject of $j(\theta)$.
\item[{\it 4 Involution:}]
The map $j$ is involutive,
i.e.: $j\circ j=\id$.
\end{description}
\eef

We will see later, that the charge conjugation is one example for a 
co-conjugation in the sense of Definition 1.3.  

%%%%%%%%%%%%%%%%%%%%%%%%%%%%%%%%%%%%%%%%%%%%%%%%%%%%%%%%%%%%%%%%%%%%%%%%%%%%%%
\section{Candidates for Antisoliton Sectors}
%%%%%%%%%%%%%%%%%%%%%%%%%%%%%%%%%%%%%%%%%%%%%%%%%%%%%%%%%%%%%%%%%%%%%%%%%%%%%
As described in the introduction, 
there are different constructions of antisoliton sectors
available. The first construction uses
the concepts of D.Guido and R.Longo \cite{GuiLo}.

For a soliton homomorphism
$\rho:a\to b$ which is localized in $W_++x$,
the PCT-conjugate homomorphism is given by
the formula $j(\rho)=J_a\rho(J_bAJ_b)J_a$, where
$J_a$ is the PCT-antiunitary with respect to the vacuum $a$.
To obtain an antisoliton homomorphism which is also localized 
in $W_++x$ one has to take 
$\bar\rho_0:=J_b\rho'(J_aAJ_a)J_b$ as an antisoliton candidate, where 
$\rho':b\to a$ is a soliton homomorphism localized in $W_--x$ 
and unitarily equivalent to $\rho:a\to b$. 
We interpret the reflection of the localization region 
as a PT-conjugation and call such a $\rho':b\to a$ a PT-conjugate for 
$\rho$. 

We apply the methods which were used by 
K.Fredenhagen \cite{Fre3} to obtain an alternative 
construction. As described in the introduction, 
one has to control the following limit:
$$
\lim_{|x|\to +\infty}\<\psi,U_\rho(x)AU_\rho(-x)\psi\>=\bar\omega_\rho(A) \ \ \
; \ \ \ \psi\in\Hc_a \ \  ; \ \  ||\psi||=1
$$
Thus we get a state 
$\bar\omega_\rho$ which belongs to a soliton sector $[\bar\rho_+]$, such that 
$[\bar\rho_+]\times[\rho]$ contains the vacuum $b$.
But there is another possibility left over. One can also take
a PT-conjugate soliton homomorphism $\rho':b\to a$,  
and show that the limit
$$
\lim_{|x|\to -\infty}\<\psi',U_{\rho'}(x)AU_{\rho'}(-x)\psi'\> =
\bar\omega_{\rho'}(A) \ \ \ ; \ \ \ \psi'\in\Hc_b \ \  ; \ \  ||\psi'||=1
$$
exists. Now, $\bar\omega_{\rho'}$ belongs to a sector 
$[\bar\rho_-]$, such that $[\rho]\times[\bar\rho_-]$
contains  the vacuum $a$.
Thus this construction leads to two further candidates
for antisoliton sectors
which are represented by the states 
$\bar\omega_\rho$ and $\bar\omega_{\rho'}$.

{\it Remark:} That one has to distinguish the two constructions of
$\bar\omega_\rho$ and $\bar\omega_{\rho'}$, is not an
effect of a difference in the vacuum sectors $a=s(\rho)$ and
$b=r(\rho)$. This problem also arises for
improper soliton homomorphisms $\rho$ with
$s(\rho)=r(\rho)$. The true reason for this distinction is the fact that
the space-like complement of a arbitrarily small
double-cone (in $d=1+1$ dimensions) is a disconnected region.

In the sequel, we 
study the question which of the following properties are 
fulfilled by each of the candidates. 

We ask, whether a particle and its antiparticle
have the same mass, i.e.:

{\it A1:} For a soliton homomorphism $\rho\in\shom(q,x)$, the
antisoliton homomorphism $\bar\rho\in\shom(q,x)$ satisfies:
$$
\sp (U_\rho)=\sp (U_{\bar\rho})
$$

We also study the question if the antiparticle carries the inverse
charge of the corresponding particle, i.e.:

{\it A2:} For a soliton homomorphism $\rho\in\shom(q,x)$
the antisoliton homomorphism $\bar\rho$ fulfills the following relations:
$$
\begin{array}{l}
s(\bar\rho)=r(\rho) \ \ \ {\rm and} \ \ \ r(\bar\rho)=s(\rho)
\vspace{0.5cm}\\
\rho\bar\rho > \id_{s(\rho)} \ \ \ {\rm and} \ \ \
\bar\rho\rho > \id_{r(\rho)}
\end{array}
$$

Furthermore, we try to find out whether
the antiparticle of the
antiparticle is the particle itself, i.e.:

{\it A3:}
For a soliton homomorphism $\rho\in\shom(q,x)$ the antisoliton
fulfills following relation:
$$
e(\ol{(\bar\rho)})=e(\rho)
$$

%%%%%%%%%%%%%%%%%%%%%%%%%%%%%%%%%%%%%%%%%%%%%%%%%%%%%%%%%%%%%%%%%%%%%%%%%%%%%
\subsection{The $C_0$-Conjugation for Soliton Sectors}
%%%%%%%%%%%%%%%%%%%%%%%%%%%%%%%%%%%%%%%%%%%%%%%%%%%%%%%%%%%%%%%%%%%%%%%%%%%%
\bdef
For a soliton homomorphism
$\rho\in\shom(q,x)$, we call a unitarily equivalent soliton
homomorphism $\rho'\in\shom(-q,-x)$ a PT-conjugate for $\rho$.
\eef
\blem
For each soliton homomorphism $\rho$ there
exists a PT-conjugate $\rho'$, unique up to inner
unitary equivalence, such that the map
$$
j_{PT}:e(\rho)\in\sec(q)\mapsto e(\rho')\in\sec(-q)
$$
is a well-defined
co-conjugation.
\elem
\bprf
According to Lemma 1.3, we conclude that $j_{PT}$ is a well-defined map. 
Since $j_{PT}$ preserves unitary equivalence, i.e.
$[j_{PT}e(\rho)]=[\rho]$, it follows that 
$j_{PT}$ is additive and isotonous.
It is also clear that $j_{PT}$ is involutive.
For $\rho_j\in\shom(q,x)$ we choose PT-
conjugates $\rho_j'\in\shom(-q,-x)$, $j=1,2$.
For $r(\rho_2)=s(\rho_1)$ we have $s(\rho_2')=r(\rho_1')$.
As mentioned in the last section, 
the representations $\rho_2'\rho_1'|_\Ag$ 
and $\rho_1\rho_2|_\Ag$ are unitarily equivalent and  
$\rho_2'\rho_1'$ is indeed a PT-conjugate for $\rho_1\rho_2$. 
Thus $j_{PT}$ fulfills the cofunctor condition.   
\eprf

For each vacuum sector $a\in\sec_0$,
the algebra $\Ag(W_q)_a$ is a von-Neumann-algebra with cyclic separating
vector $\Omega_a$. 
The modular data of $(\Ag(W_+)_a,\Omega_a)$ are denoted by
$(J_a,\Delta_a)$.
According to the theorem of Borchers \cite{Bor1}, we obtain the relation
$$
J_aU_a(x)J_a=U_a(-x)
$$
for each $x\in\Rl^2$ and for each vacuum sector $a$.
We define $j_a:=\ad(J_a)$, which maps $\Ag_a^+$ onto
$\Ag_a^-$ and vice versa.
For a soliton homomorphism $\rho\in\shom(q,x)$, we define
the *-homomorphism 
$$
j(\rho):=j_{s(\rho)}\circ\rho\circ j_{r(\rho)}(A)
$$
which maps
$\Ag_{r(\rho)}^{-q}$ into $\Ag_{s(\rho)}^{-q}$.
\bass
Let us assume that there exists PCT-symmetry, i.e. 
an involutive antiautomorphism 
$j:\Ag\mapsto\Ag$ with $j(\Ag(\Oc))=\Ag(-\Oc)$ 
and $j\circ\alpha_x=\alpha_{-x}\circ j$ which is 
implemented in each vacuum sector $a\in\sec_0$ by the
modular conjugation $J_a$, i.e.:
$$
\id_a(jA)=J_a\id_a(A)J_a
$$
\eass    
\bpro
Let $j(\rho)=j_{s(\rho)}\circ\rho\circ j_{r(\rho)}$
be the *-homomorphism defined as above.
If the soliton homomorphism $\rho$ is a
one soliton homomorphism, then
the *-homomorphism $j(\rho)$
is also a one soliton homomorphism which  
has the same mass spectrum as $\rho$.
If $\rho$ is a *-homomorphism 
localized in $W_q+x$, then $j(\rho)$ is localized in
$W_{-q}-x$.
\epro
\bprf
It is not hard to check the following equation:
$$
j(\rho)(\alpha_xA)= J_{s(\rho)}U_\rho(-x)J_{s(\rho)}j(\rho)(A)
J_{s(\rho)}U_\rho(x)J_{s(\rho)}  \ \  ; \ \ A\in\Ag
$$
Thus, $U_{j(\rho)}(x):=J_{s(\rho)}U_\rho(-x)J_{s(\rho)}$
implements the translation group in the
representation $j(\rho)|_\Ag$.
Since $\sp(U_\rho)=\sp(U_{j(\rho)})$ and $\rho$ is a massive
one-particle representation, $j(\rho)$ is also a
massive one-particle representation which
has the same mass spectrum as $\rho$.
Now, for each $A\in\Ag(\Oc)$ with $\Oc\subset W_q+x$ we obtain
$$
\begin{array}{l}
j_{s(\rho)}\circ\rho\circ j_{r(\rho)}(A)=
j_{s(\rho)}\circ\rho\circ \id_{r(\rho)}(jA)\vspace{0.2cm}\\
=j_{s(\rho)}\circ \id_{s(\rho)}(jA)=\id_{s(\rho)}(jjA)\vspace{0.2cm}\\
=\id_{s(\rho)}(A)
\end{array}
$$
Thus $j(\rho)$ is localized in $W_{-q}+x$.
\eprf
\btho

\begin{description}
\item[{\it i:}]
The map $j_{PCT}:\sec(q)\to \sec(-q)$ which is defined by
$$
j_{PCT}e(\rho):=e(j(\rho))
$$
is a conjugation (Definition 1.3), called
PCT-conjugation.
\item[{\it ii:}]
The PCT-conjugation commutes with the
PT-conjugation, i.e.:
$$
j_0:=j_{PCT}\circ j_{PT}
=  j_{PT} \circ j_{PCT}:\sec(q)\to\sec(q)
$$
The co-conjugation
$j_0$ is called $C_0$-conjugation and
for a soliton homomorphism $\rho\in\shom(q,x)$ we call 
$\bar\rho_0\in\shom(q,x)$ with $e(\bar\rho_0)=j_0e(\rho)$
a $C_0$-conjugate for $\rho$.
\end{description}
\etho
\bprf
{\it i:}
By construction, the
soliton homomorphism $j(\rho)$ is contained in 
the set $\shom(-q,-x)$,
if $\rho$ belongs to $\shom(q,x)$.
Thus the map $j$ is a well defined function.
For $\rho_j\in\shom(q,x_j)$, $j=1,2$, $s(\rho_2)=r(\rho_1)$,
we have:
$$
\begin{array}{l}
j(\rho_1\rho_2)=
j_{s(\rho_1)}\circ\rho_1\rho_2\circ j_{r(\rho_2)}
\vspace{0.2cm}\\
=j_{s(\rho_1)}\circ\rho_1\circ j_{s(\rho_1)}\circ j_{s(\rho_1)}\circ
\rho_2\circ j_{r(\rho_2)}=j(\rho_1)j(\rho_2)
\end{array}
$$
Hence $j$ is a functor.  
Let $\rho,\hat\rho$ be soliton homomorphisms in $\shom(q,x)$
and $w\in I(\rho,\hat\rho)$ an intertwiner.
It is easy to see that $J_{s(\rho)}w J_{s(\rho)}$ intertwines
$j(\rho)$ and $j(\hat\rho)$. From this we conclude
that $j$ is additive and isotonous. Moreover, we have $j\circ j=\id$ and hence
$j$ is a conjugation. Since $j$ preserves inner unitary equivalence,
we conclude that $j_{PCT}$ is also a well defined conjugation.    

{\it ii:}
To prove {\it ii} we have to show the following relation:
For $\rho_1\in\shom(q,x)$ and $\rho_2\in\shom(-q,y)$:
\begin{equation}
[\rho_1]=[\rho_2] \ \ \ra \ \ [j(\rho_1)] = [j(\rho_2)]
\end{equation}
From this we obtain for $\rho\in\shom(q,x)$ that 
$$
e(j(\rho))=j_{PT}e(j(\rho')) \ \ ,
$$
where $\rho'$ is a PT-conjugate
for $\rho$.
Remember that $j(\rho)$ and $\rho'$ are both contained in $\shom(-q,-x)$.
Thus we have
$$
j_{PCT}j_{PT}e(\rho)
=j_{PT}j_{PCT}e(\rho)
$$
and {\it ii} is proven.

It remains to prove the relation (2) above.
For equivalent $\rho_1\in\shom(q,x)$ and
$\rho_2\in\shom(-q,y)$
we have $a=s(\rho_1)=r(\rho_2)$ and $b=r(\rho_1)=s(\rho_2)$.
Furthermore, there is a unitary operator
$u:\Hc_b\to\Hc_a$ which intertwines
$\rho_1|_\Ag$ and $\rho_2|_\Ag$.
In the following, we show that $u^J:=J_auJ_b$
intertwines the representations $j(\rho_1)|_\Ag$ and
$j(\rho_2)|_\Ag$.
Let $\Oc$ be an arbitrary double cone, then for any $A\in\Ag(\Oc)$
we obtain:
$$
j(\rho_1)(A)u^J=J_a\rho_1(jA)uJ_b
=J_au\rho_1(jA)J_b=J_auJ_bJ_b\rho_2(jA)J_b
=u^Jj(\rho_2)(A)
$$
Thus $j(\rho_1)$ and $j(\rho_2)$ are unitarily equivalent.
\eprf

Using the statements of Theorem 2.1, we 
conclude this subsection and summarize the properties of a 
$C_0$-conjugate in the following corollary:
\bcor
Suppose $\theta$ is a one-soliton sector,
then the $C_0$-conjugate 
$j_0\theta$ is also a one-soliton sector with the same
mass spectrum as $\theta$ (property {\it A1}). Moreover, 
the $C_0$-conjugate of the $C_0$-conjugate of 
an inner soliton sector $\theta$ is 
$\theta$ itself (property {\it A3}).
\ecor

%%%%%%%%%%%%%%%%%%%%%%%%%%%%%%%%%%%%%%%%%%%%%%%%%%%%%%%%%%%%%%%%%%%%%%%%%%%%%%
\subsection{The $C_\pm$-Conjugation for Soliton Sectors}
%%%%%%%%%%%%%%%%%%%%%%%%%%%%%%%%%%%%%%%%%%%%%%%%%%%%%%%%%%%%%%%%%%%%%%%%%%%%%%
Beside the $C_0$-conjugate for a one soliton homomorphism
there are two further candidates for antisoliton sectors
which can be represented by *-homomorphisms of
soliton states.
We start this subsection by presenting the main result to
establish the existence of these *-homomorphisms.
Afterwards, we apply the results in \cite{Fre3} to prepare the proof.

For convenience,
it is sufficient to formulate the theorem for soliton homomorphisms
with orientation $q=+$. The formulation for the 
case $q=-$ can be done in complete analogy.  
%%%%%%%%%%%%%%%%%%%%%%%%%%%%%%%%%%%%%%%%%%%%%%%%%%%%%%%%%%%%%%%%%%%%%%%%%%%%
\btho
Let $\rho:a\to b\in\shom_1(+,x)$ be a one-soliton homomorphism.
Then there are irreducible soliton homomorphisms
$\bar\rho_\pm:b\to a\in\shom(+,x)$, let us call them 
$C_\pm$-conjugates for $\rho:a\to b$,
with the following properties:
\begin{description}
\item[{\it i:}]
The energy momentum spectrum of $U_{\bar\rho_\pm}$ is
contained in the closed forward light cone, i.e.:
$\sp (U_{\bar\rho_\pm})\subset \bar V^+$.
\item[{\it ii:}]
The representation $\bar\rho_+\rho$ contains 
$\id_b$ precisely once and the
representation $\rho\bar\rho_-$ contains 
$\id_a$ precisely once.
%\item[iii:]
%The properties {\it ii} determine the inner unitary equivalence classes of
%$\bar\rho_+$ and $\bar\rho_-$. 
\end{description}
\etho
%%%%%%%%%%%%%%%%%%%%%%%%%%%%%%%%%%%%%%%%%%%%%%%%%%%%%%%%%%%%%%%%%%%%%%%%%%%%%

Let $\rho$ a massive one soliton homomorphism and
let $\bar\rho_\pm$ $C_\pm$- conjugates for $\rho$.
Since $\bar\rho_{1,\pm}$ and $\bar\rho_\pm$ are inner unitarily equivalent,
if $\rho$ and $\rho_1$ are inner unitarily equivalent, we may  
define the {\sl $C_\pm$-conjugation} $j_\pm$ by
$$
j_\pm:e(\rho)\in \sec_1(q)\mapsto e(\bar\rho_\pm)\in\sec(q)  \ \ \ .
$$

{\it Remark:} From Theorem 2.2, one cannot
decide whether $\bar\rho_\pm$ are massive
one-particle representation. In addition, there is
no reason why $\bar\rho_+$ and $\bar\rho_-$ should be
equivalent.
Hence Theorem 2.2 does not imply that 
$j_\pm$ are co-conjugations in the sense of Definition 3.1.

To prove Theorem 2.2, we apply the results 
and methods used in \cite{Fre3}.    
%%%%%%%%%%%%%%%%%%%%%%%%%%%%%%%%%%%%%%%%%%%%%%%%%%%%%%%%%%%%%%%%%%%%%%%%%%%%%%
\bpro
If $\rho:a\to b\in\shom_1(+,x)$ and $\rho':b\to a\in\shom_1(-,-x)$
are PT-conjugate one soliton homomorphisms, then 
there are translation covariant states, $\bar\omega_\rho:\Ag^+_a\to\Cl$
and
$\bar\omega_{\rho'}:\Ag_b^-\to\Cl$
with the following properties:
\begin{description}
\item[{\it 1:}]
Let $e_\pm\in W_\pm$ be any space-like vector of unit length.
Then, for all $A\in\Ag_a^+$ one has
$$
\lim_{r\to\infty}
\<\psi,U_\rho(re_+)AU_\rho(re_+)^*\psi\>=\bar\omega_\rho(A)
\ \ \ ; \ \ \ \forall \psi\in\Hc_a \ \ \ ; \ \ \ ||\psi||=1
$$
and for all $A\in\Ag_b^-$ one obtains
$$
\lim_{r\to\infty}
\<\psi',U_{\rho'}(re_-)AU_{\rho'}(re_-)^*\psi'\>
=\bar\omega_{\rho'}(A)
\ \ \ ; \ \ \ \forall \psi'\in\Hc_b \ \ \ ; \ \ \ ||\psi'||=1
$$
\item[{\it 2:}]
The state $\bar\omega_\rho$ is 
normal on the wedge algebras $\Ag(W_++x)_a$, the
similar statement holds also for $\bar\omega_{\rho'}$.
\item[{\it 3:}]
For soliton homomorphisms $\rho_1\in\shom(+,y)$ inner
unitarily equivalent to $\rho$, 
$\bar\omega_{\rho_1}$ and $\bar\omega_\rho$ are equivalent states.
An analogous statement holds for a pair of inner equivalent
one soliton homomorphisms with orientation $q=-$.
\item[{\it 4:}]
The GNS-representations $(\bar\Hc_+,\bar\pi_+,\xi_+)$ of $\omega_\rho$ and 
$(\bar\Hc_-,\bar\pi_-,\xi_-)$ of $\omega_{\rho'}$ 
can be localized in any wedge 
region, i.e. there are  
unitary operators $v_{\pm,a}:\bar\Hc_\pm\to\Hc_a$ and 
$v_{\pm,b}:\bar\Hc_\pm\to\Hc_b$
such that for $A\in\Ag(W_+)$ and $A'\in\Ag(W_-)$ the equations
$$
\bar\pi_\pm(A)=v_{\pm,a}^*\id_a(A)v_{\pm,a}  \ \ \ {\rm and} \ \ \
\bar\pi_\pm(A')=v_{\pm,b}^*\id_b(A')v_{\pm,b} 
$$
hold.
\end{description}
\epro
\bprf 
One can prove the proposition
with the same methods used in the proof of 
Lemma 2.2 of ref. \cite{Fre3} (p. 144-146). 
Let us sketch only the main ideas. 

Applying the results of \cite{BuFre}, for a neighborhood
$\Delta\subset\Rl^2$ of the mass shell with
$\Delta \cap \sp(U_\rho)\subset H_m$, there is a quasi local
operator $B\in\Ag^+_b$, such that $\Psi_\Delta:=\rho(B)E_\rho(\Delta)\not=0$.
Here $E_\rho$ denotes the spectral measure with respect to 
the translation group $U_\rho$.  
Now one can show, that there are states
$$
\varphi_\pm:\ol{\cup_x\rho(\Ag(W_\mp+x))'}^{||\cdot ||}\to \Cl
$$
such that for each $\psi_\Delta=\Psi_\Delta\psi$, $\psi\in\Hc_a$
$$
r\mapsto 
\<\psi_\Delta,(U_\rho(re_\pm)AU_\rho(-e_\pm)-\varphi_\pm(A)\1)\psi_\Delta\>
$$
is a function of fast decrease. 
Moreover, $\varphi_\pm$ is a limit 
of states
$$
\varphi_\pm^r(A):=
\<\psi_\Delta,(U_\rho(re_\pm)AU_\rho(-e_\pm)\psi_\Delta\>
$$
which are normal on $\rho(\Ag(W_\mp+x))'$ and hence
$\varphi_\pm$ is normal on $\rho(\Ag(W_\mp+x))'$.
If $\rho$ is localized in $W_++x$, then one gets 
$$
\ol{\cup_x\rho(\Ag(W_-+x))'}^{||\cdot ||}=\Ag_a^+
$$
and we obtain the state $\bar\omega_\rho:=\varphi_+:\Ag_a^+\to\Cl$.
On the other hand, if $\rho':=\rho$ is localized in 
$W_-+x$, then one gets
$$
\ol{\cup_x\rho'(\Ag(W_-+x))'}^{||\cdot ||}=\Ag_b^-
$$  
and we obtain the state $\bar\omega_{\rho'}:=\varphi_-:\Ag_b^-\to\Cl$.  

By construction, the state $\bar\omega_\rho$ is invariant under 
the charged translations, i.e.: $\bar\omega_\rho(U_\rho(x)AU_\rho(-x))=
\bar\omega_\rho(A)$. Thus one obtains via
$$
T_+(x)(\bar\pi_+(A)\xi_+):=\bar\pi_+(U_\rho(x)A)\xi_+
$$
a strongly continuous representation of the translation group.
Applying the methods of \cite{Fre3}, it can be shown that the  
spectrum of $T_+$ is contained in the closed forward light cone and 
that the vector $\xi_+$ is a ground-state of $T_+$, for the simple 
eigenvalue $0$. Thus   
one concludes that $\bar\pi_+$ is irreducible. 
We define now another unitary representation of the translation 
group by
$$
x\mapsto U_{\bar\pi_+}(x):=\bar\pi_+(U_a(x)U_\rho(-x))T_+(x)  \ \
$$
which implements $\alpha_x$ in the representation $\bar\pi_+$.
It can be shown that $U_{\bar\pi_+}$ is strongly continuous.
The analogous result is also true for $\bar\omega_{\rho'}$ and 
we have established {\it 1} and {\it 2}.

Let $\rho_1$ be a soliton homomorphism which is inner equivalent to $\rho$.
Then one computes that $\bar\omega_\rho= \bar\omega_{\rho_1}\circ\ad(u)$
holds, where $u$ is a unitary intertwiner from $\rho_1$ to $\rho$.
Thus {\it 3} is proven.

By construction, one has $\bar\omega_\rho\circ\rho=\omega_b$.     
If $\rho$ is localized in $W_++x$, one obtains for $A'\in\Ag(W_-+x)$:
$$
\bar\omega_\rho(A')=\bar\omega_\rho\circ\rho(A')=\omega_b(A') \ \ \ .
$$ 
Hence there is a unitary operator $v_{+,b}$, such that for each 
$A'\in\Ag(W_-+x)$ holds the equation:  
$$
v_{+,b}\bar\pi_+(A')=\id_b(A')v_{+,b}    
$$
Since $\bar\omega_\rho$ is normal on $\Ag(W_++x)_a$, there is a 
vector $\phi\in\Hc_a$, such that 
$$
\bar\omega_\rho(A):=\<\phi,\id_a(A)\phi\> 
$$
for each $A\in\Ag(W_++x)$. Now we define an isometry $w_1$ by
$$
w_1(\bar\pi_+(A)\xi_+):=\id_a(A)\phi \ \ \ ; \ \ \ A\in\Ag(W_++x)_a
$$
and obtain that $E_1:=w_1w_1^*$ is contained in $\Ag(W_-+x)_a$.
According to \cite{Bor3} there is an 
isometry $w_a\in\Ag(W_-+x)_a$ such that
$w_aw_a^*=E_1$
and for the unitary $v_{+,a}:=w_a^*w_1$ holds the equation:
$$
v_{+,a}\bar\pi_+(A)=\id_a(A)v_{+,a} 
$$
The analogue can also be proven for $\bar\omega_{\rho'}$ and one gets {\it 4}.
\eprf
%%%%%%%%%%%%%%%%%%%%%%%%%%%%%%%%%%%%%%%%%%%%%%%%%%%%%%%%%%%%%%%%%%%%%%%%%%%%
%%%%%%%%%%%%%%%%%%%%%%%%%%%%%%%%%%%%%%%%%%%%%%%%%%%%%%%%%%%%%%%%%%%%%%%%%

%Before we prove Theorem 2.2, we show that
%the $C_\pm$-conjugates 
%of a massive one soliton homomorphism $\rho$
%are fixed by its properties uniquely up to inner
%unitary equivalence. 
To prove Theorem 2.2,
it is sufficient to consider the case that $\rho$ has 
orientation $q=+$. 
%%%%%%%%%%%%%%%%%%%%%%%%%%%%%%%%%%%%%%%%%%%%%%%%%%%%%%%%%%%%%%%%%%%%%%%%%
%\blem
%Let $\rho\in\shom_1(+,x)$ be a massive one-soliton homomorphism.
%If there exists a pair $\sgm_\pm\in\shom_1(+,x)$
%of massive one-soliton homomorphisms
%with $\sgm_+\rho>\id_{r(\rho)}$ and $\rho\sgm_->\id_{s(\rho)}$,
%then we have:
%$$
%e(\sgm_\pm)=j_\pm e(\rho)
%$$
%\elem
%\bprf
%Let $\bar\rho_\pm$ be any representative of $j_\pm e(\rho)$.
%With the notation above, there are isometries
%$v,w$, such that
%$$
%\bar\rho_+\rho(A)v=  vA  \ \ \ {\rm and} \ \ \
%\sgm_+\rho(A)w=  wA  \ \ \ .
%$$
%Thus we obtain {\sl normal conditional expectations} 
%$$
%\epsilon_{\rho,1}=\rho(v^*\bar\rho_q(\cdot)v)
%\ \  {\rm and} \ \
%\epsilon_{\rho,2}=\rho(w^*\sgm_q(\cdot)w)
%$$
%from $\Ag(W_q+x)_{s(\rho)}$ to $\rho(\Ag(W_q+x)_{r(\rho)}$.
%Since $\rho$ is irreducible, there exists only one
%such conditional expectation.
%Thus one conclude for the leftinverse $\phi_\rho$ of $\rho$:
%$$
%\phi_\rho(A)=v^*\bar\rho_+(A)v=w^*\sgm_+(A)w
%$$
%Hence both $\sgm_+$ and $\bar\rho_+$ are conjugates with respect to the 
%leftinverse $\phi_\rho$ and therefore inner unitarily equivalent.
%\eprf
%%%%%%%%%%%%%%%%%%%%%%%%%%%%%%%%%%%%%%%%%%%%%%%%%%%%%%%%%%%%%%%%%%%%%%%%%%%%%
\bprf (Theorem 2.2:)
{\it i:}
By Proposition 2.2, we obtain that $\bar\omega_\rho$ and $\bar\omega_{\rho'}$
are translation covariant states, and if we use Proposition 2.2.{\it 4}
it follows that $\bar\omega_\rho$ and $\bar\omega_{\rho'}$ are soliton states.
Since the spectra of the translation groups $T_\pm$, 
$U_a$ and $U_\rho$ are contained in the closed 
forward light cone, one obtains by applying the 
additivity of the energy momentum spectrum (see \cite{Fre3,DHRI,DHRII})
that the spectrum of $U_{\bar\pi_\pm}$ is also contained in the closed 
forward light cone.

{\it ii:}
The statement {\it ii} follows immediately from the construction of the 
states $\bar\omega_\rho$ and $\bar\omega_{\rho'}$.
\eprf
%{\it iii:}
%The statement {\it iii} is a direct consequence of Lemma 2.4.
%\eprf

%%%%%%%%%%%%%%%%%%%%%%%%%%%%%%%%%%%%%%%%%%%%%%%%%%%%%%%%%%%%%%%%%%%%%%%%%%%%%
\section{On the Existence of Antisoliton Sectors}
%%%%%%%%%%%%%%%%%%%%%%%%%%%%%%%%%%%%%%%%%%%%%%%%%%%%%%%%%%%%%%%%%%%%%%%%%%%%%
The results of the last section shows that a
$C_0$-conjugate $\bar\rho_0$ for a 
one soliton homomorphism $\rho$ has the property 
{\it A1}, i.e.
$\bar\rho_0$ is a one soliton homomorphism having the same mass
spectrum as $\rho$, and property
{\it A3}, i.e. a
$C_0$-conjugate of $\bar\rho_0$ is inner
unitarily equivalent to $\rho$.
For the $C_-$-conjugate $\bar\rho_-$ one can show the
first part of {\it A2}, i.e.
$\rho\bar\rho_-$ contains the identity $\id_{s(\rho)}$
precisely once. The $C_+$-conjugate $\bar\rho_+$
satisfies the second part of {\it A2}, i.e.
$\bar\rho_+\rho$ contains the identity $\id_{r(\rho)}$
precisely once.

As described in the introduction, the square of the
dimension of a soliton homomorphism $\rho$ is defined by the
index of the inclusion
$\rho(\Ag(W_q+x)_{r(\rho)})\subset\Ag(W_q+x)_{s(\rho)}$, i.e.:
$$
d(\rho)^2=\Ind(\rho(\Ag(W_q+x)_{r(\rho)}),\Ag(W_q+x)_{s(\rho)})
$$
Here $q$ is the orientation of $\rho$ and $W_q+x$ its localization
region.

We show
that the index of an inner soliton sector $\theta$ is finite, if and
only if there exist an inner soliton sector $\bar\theta$, such that
$\bar\theta\theta$ contains the identity $i_{r(\theta)}$ and
$\theta\bar\theta$ contains the identity $i_{s(\theta)}$. 
As mentioned in the introduction, this is a slight 
generalization of Longo's result \cite{Lo2} (Theorem 4.1) and can be 
proven with similar methods. 
To make arguments clear, we shall give here an explicit proof.
   
Then one can use this statement to prove the main result 
of this section which states that for a soliton sector 
$\theta$ there is
an antisoliton sector $\bar\theta$ with the 
properties {\it A1}- {\it A3}, if one of the following three
equivalent conditions is fulfilled:
\begin{description}
\item[{\it 1:}]
The index or dimension of $\theta$ is finite.
\item[{\it 2:}]
The $C_+$-conjugate and the $C_-$-conjugate of
$\theta$ are equal.
\item[{\it 3:}]
Either the $C_+$- or the $C_-$-conjugate equals the $C_0$-conjugate.
\end{description}

We have to mention that there might be a one soliton sector with
infinite dimension. In such a case there is no unique
choice for an antisoliton homomorphism.

%%%%%%%%%%%%%%%%%%%%%%%%%%%%%%%%%%%%%%%%%%%%%%%%%%%%%%%%%%%%%%%%%%%%%%%%%%%%%%%%%%
\subsection{Relation between $C_\pm$-Conjugate and
$C_0$-Conjugate Homomorphisms}
%%%%%%%%%%%%%%%%%%%%%%%%%%%%%%%%%%%%%%%%%%%%%%%%%%%%%%%%%%%%%%%%%%%%%%%%%%%%%
One obtains the following nice and useful relation between the
$C_\pm$- and the $C_0$-conjugation.
\btho
The equation
$$
j_0\circ j_+=j_-\circ j_0:\sec_1(q)\to
\sec(q)
$$
holds on the set of all inner one soliton sectors.
\etho
\bprf
Let $\rho_q$, $q=\pm$, be a pair of PT-conjugate massive one
soliton homomorphisms. Let us choose
representatives $\bar\rho_{qp}$ of $j_p(e(\rho_q))$; $p,q=\pm$, where
$\bar\rho_{qp}$ is localized in the same region as $\rho_q$,
i.e. $W_q+qx$.

We choose now a vector $\psi$ with $||\psi||=1$ which is contained in 
$\Pc_{s(\rho_+)}:=\ol{\{Aj_{s(\rho_+)}(A)\Omega_{s(\rho_+)}\}}$.
For $\psi$ we have $J_{s(\rho_+)}\psi=\psi$ and
thus we obtain for each observable $A\in\Ag$:
$$
\lim_{|y|\to -\infty}
\<\psi,U_{j(\rho_+)}(y)AU_{j(\rho_+)}(-y)\psi\>
=\bar\omega_{j(\rho_+)}(A)
$$
Since $j(\rho_+)$ is a massive one soliton homomorphism, localized
in $W_--x$, the limit above
exists. Hence
$\ol{j(\rho_-)}_-$ is a representative of $j_-e(j(\rho_-))$ and
we obtain:
$$
[\bar\omega_{j(\rho_+)}]
=[\omega_{r(\rho_+)}\circ \ol{j(\rho_-)}_-]
$$
On the other hand we have:
$$
\begin{array}{l}
\lim_{|y|\to -\infty}
\<\psi,U_{j(\rho_+)}(y)AU_{j(\rho_+)}(-y)\psi\>\vspace{0.2cm}\\
=
\lim_{|y|\to -\infty}
\<\psi,J_{s(\rho_+)}U_{j(\rho_+)}(y)A
U_{j(\rho_+)}(-y)J_{s(\rho_+)}\psi\>
\vspace{0.2cm}\\
=
\lim_{|x|\to +\infty}
\<\psi,U_{\rho_+}(x)J_{s(\rho_+)}AJ_{s(\rho_+)}
U_{\rho_+}(-x)\psi\>\vspace{0.2cm}\\
=
\bar\omega_{\rho_+}(J_{s(\rho_+)}AJ_{s(\rho_+)})
\end{array}
$$
The state $\bar\omega_{j(\rho_+)}$ is equivalent to
$\omega_{r(\rho_+)}\circ j(\bar\rho_{++})$.
Hence we obtain the equation:
$$
[j(\bar\rho_{++})]=[\ol{j(\rho_-)}_-]
$$
Since $j(\bar\rho_{++})$ is localized in $W_--x$ and
$\ol{j(\rho_-)}_-$ is localized in $W_++x$ and $j(\bar\rho_{++})$ and
$\ol{j(\rho_-)}_-$ are unitarily equivalent, we obtain
$$
\begin{array}{l}
j_0j_+e(\rho_+) = j_{PT}j_{PCT}e(\bar\rho_{++})
=j_{PT}e(j(\bar\rho_{++})
\vspace{0.2cm}\\
=e(\ol{j(\rho_-)}_-)
=j_-e(j(\rho_-))=j_-j_0e(\rho_+)
\end{array}
$$
which completes the proof.
\eprf

%%%%%%%%%%%%%%%%%%%%%%%%%%%%%%%%%%%%%%%%%%%%%%%%%%%%%%%%%%%%%%%%%%%%%%%%%%%%
\subsection{The Dimension for Soliton Homomorphisms}
%%%%%%%%%%%%%%%%%%%%%%%%%%%%%%%%%%%%%%%%%%%%%%%%%%%%%%%%%%%%%%%%%%%%%%%%%%%%%
%As mentioned at the begin of this section, we define a dimension
%function on the set of soliton sectors. We will see later that this is
%a nice tool to obtain further information about
%the three antisoliton candidates.

Let $\rho\in\shom(q,x)$ be a *-homomorphism.
Then, we define the square of the dimension of $\rho$, by
$$
d(\rho)^2:=\Ind(\rho):=\Ind(\rho(\Ag(W_q+x)_{r(\rho)}),\Ag(W_q+x)_{s(\rho)})
$$
Since the dimension of an inner soliton sector depends 
only on the inner equivalence class
of a *-homomorphism, one can define the dimension of an 
inner soliton sector 
$\theta=e(\rho)\in\sec(q)$ by $d(\theta):=d(\rho)$.

A consequence of the results proven in
\cite{Lo1,Lo2,Lo4,Lo5,KoLo}
is the following proposition:
\bpro
All *-homomorphisms $\rho_1,\rho_2\in\shom(q,x)$
with finite dimension have the following properties:
\begin{description}
\item[{\it i Multiplicativity:}]
For $s(\rho_2)=r(\rho_1)$:
$$
d(\rho_1\rho_2)=d(\rho_1)d(\rho_2)
$$
\item[{\it ii Additivity:}]
For $s(\rho_1)=s(\rho_2)$ and $r(\rho_1)=r(\rho_2)$ and
one has:
$$
d(\rho_1\oplus\rho_2)=d(\rho_1)+d(\rho_2)
$$
\item[{\it iii Reducibility:}]
If $s(\rho_2)=r(\rho_1)$, then there are finitely many 
soliton homomorphisms $\sgm_j$ with $s(\sgm_j)=s(\rho_2)$ 
and $r(\sgm_j)=r(\rho_1)$, such that
$$
\rho_1\rho_2=\bigoplus_j N_{\rho_1,\rho_2}^j \sgm_j
$$
where $N_{\rho_1,\rho_2}^j$ are natural numbers.   
\end{description}
\epro

%%%%%%%%%%%%%%%%%%%%%%%%%%%%%%%%%%%%%%%%%%%%%%%%%%%%%%%%%%%%%%%%%%%%%%%%%%%%
\subsection{Conditions for the Existence of Antisolitons}
%%%%%%%%%%%%%%%%%%%%%%%%%%%%%%%%%%%%%%%%%%%%%%%%%%%%%%%%%%%%%%%%%%%%%%%%%%%%
We come now to the main result of this section:
\btho
For every massive inner one soliton sector $\theta\in\sec_1(q)$
the following statements are equivalent:
\begin{description}
\item[{\it i:}]
The dimension of $\theta$ is finite:
$$
d(\theta)<\infty
$$
\item[{\it ii:}]
The $C_+$-conjugate is equal to the $C_-$-conjugate
of $\theta$:
$$
j_+\theta=j_-\theta
$$
\item[{\it iii:}]
The $C_0$-conjugate is equal to the $C_+$- {\sl or} the
$C_-$-conjugate:
$$
j_0(\theta)=j_q(\theta) \ \ \ ; \ \ q=\pm
$$
\end{description}
\etho

We shall need some abstract index theoretical results to prove
Theorem 3.2, and before we present them, we sketch the basic 
ideas of the proof.

We prove, if statement {\it i} holds for $\theta$ then there exists an
inner soliton sector $\bar\theta$ such that $\theta\bar\theta$
contains the identity $i_{s(\theta)}$ and $\bar\theta\theta$
contains the identity $i_{r(\theta)}$. Such an inner soliton sector
exists if and only if the index of $\theta$ is finite.
Since $j_\pm\theta$
are fixed uniquely by their properties, i.e.
$j_+\theta\theta$ contains $i_{r(\theta)}$ and
$\theta j_-\theta$ contains $i_{s(\theta)}$, we obtain
$j_+\theta=j_-\theta=\bar\theta$ which implies {\it ii} and {\it iii}.
Using the relation $j_+j_0\theta=j_0j_-\theta$, 
we obtain from {\it iii} $j_\pm=j_0$ and we conclude
that the index of $\theta$ is finite.

The basic ingredient for the proof of Theorem 3.2 is the following
Lemma:
\blem
The following two statements are equivalent:
\begin{description}
\item[{\it 1:}]
The dimension of an inner sector $\theta\in\sec_1(q)$
is finite, i.e. $d(\theta)<\infty$.
\item[{\it 2:}]
There exists an inner sector $\theta'\in\sec(q)$ with
$s(\theta)=r(\theta')$ and $r(\theta)=s(\theta')$, such that
$\theta \theta'$ contains the
identity $i_{s(\theta)}$ precisely once {\sl and}
$\theta'\theta$ contains the identity
$i_{r(\theta)}$ precisely once, i.e.:
$$
\theta\theta' > i_{r(\theta)}
\ \ \ {\rm and} \ \ \ 
\theta'\theta > i_{s(\theta)}
$$
\end{description}
Furthermore, if $\theta'$ exists, then $\theta'=j_0(\theta)$.
\elem
To prove the lemma, we need two results proven in \cite{Lo2}:
\bpro
Let $\Ng\subset\Mg$ be a standard inclusion of infinite factors and
denote by $C(\Mg,\Ng)$ the collection of all
normal conditional expectations $\epsilon:\Mg\to\Ng$.
Furthermore, let $\Hc(\gamma,\Mg)$ be the Hilbert space space of all
intertwiners $\gamma(A)v=vA$, $A\in\Mg$, where $\gamma$ is a normal
endomorphism of $\Mg$.
Under this conditions, the maps
$$
\begin{array}{l}
\epsilon:\Hc(\gamma,\Ng)_1\to C(\Mg,\Ng) \ \ ; \ \ 
v\mapsto  \epsilon_v=v^*\gamma(\cdot)v
\vspace{0.2cm}\\
\epsilon^1:\Hc(\gamma,\Mg)_1\to C(\Mg_1,\Mg) \ \ ; \ \
w\mapsto \epsilon_w^1=w^*\gamma(\cdot)w
\end{array}
$$
are bijective. Here $\Hc(\gamma,\Mg)_1$ denotes
the unit sphere in $\Hc(\gamma,\Mg)$ and $\Mg_1$ denotes
the Jones basic construction, i.e. $\Mg_1=J_{\Mg}\Ng'J_{\Mg}$.
\epro
\bpro
Let $\Ng\subset\Mg$ an inclusion of infinite von-Neumann-algebras,
then the following statements are equivalent:
\begin{description}
\item[{\it 1:}]
The index of the inclusion $\Ng,\Mg$ is finite:
$$
\Ind(\Ng,\Mg)<\infty
$$
\item[{\it 2:}]
The relative commutant $\Ng'\cap\Mg$ is a finite dimensional
algebra {\sl and} there exists a faithful conditional
expectation $\epsilon:\Mg\to\Ng$ {\sl and} there
exists a faithful conditional expectation
$\epsilon':\Ng'\to\Mg'$.
\end{description}
\epro
\bprf (Lemma 3.1:)
First, let us remark that the relative commutant
$\rho(\Mg_{r(\rho)})'\cap\Mg_{s(\rho)}=
\Cl\cdot \1_{s(\rho)}$ is trivial, because $\rho$ is irreducible.
{\it 1}$\ra${\it 2}:
Suppose the index of $\theta$ is finite and
choose a representative $\rho$ of $\theta$.
Then there exists a faithful conditional expectation
$\epsilon\in C(\Mg_{s(\rho)},\rho(\Mg_{r(\rho)}))$ and
a faithful conditional expectation $\epsilon'\in
C(\rho(\Mg_{r(\rho)})',\Mg_{s(\rho)}')$.
There is also a faithful conditional
expectation $\epsilon^1\in C(\Mg_{s(\rho),1},\Mg_{s(\rho)})$ and
$\epsilon^1$ can be written in the form:
$$
\epsilon^1=\bar v^*\rho\bar\rho(\cdot)\bar v
$$
with an isometry $\bar v\in\Hc(\rho\bar\rho,\Mg_{s(\rho)})$.
Here $\bar\rho$ is any representative of $j_0(\theta)$.
Now, $\bar v$ is an intertwiner from $\rho\bar\rho$ to
$\id_{s(\rho)}$, and since $\rho$ is irreducible we
obtain that $\theta j_0(\theta)$ contains the identity
precisely once.
On the other hand, $\epsilon$ can be written in the form:
$$
\epsilon=\rho(v^*)\rho\bar\rho(\cdot)\rho(v)
$$
with an isometry $\rho(v)\in \Hc(\rho\bar\rho,\rho(\Mg_{r(\rho)}))$.
Now, $v$ intertwines $\bar\rho\rho$ and $\id_{r(\rho)}$, and
since $\rho$ is irreducible we conclude, that
$j_0(\theta)\theta$ contains the identity precisely once.

{\it 2}$\ra${\it 1}:
Suppose, there is an inner soliton sector $\theta'$, such
that $\theta\theta'> i_{s(\theta)}$ and
$\theta'\theta > i_{r(\theta)}$.
If we choose representatives $\rho$ of $\theta$ and $\rho'$ of $\theta'$,
then there are isometries $v\in\Mg_{r(\rho)}$, $\bar v\in
\Mg_{s(\rho)}$, such that:
$$
\begin{array}{l}
\rho\rho'(A)\bar v=\bar v A \ \ ; \ \ A\in\Mg_{s(\rho)}
\vspace{0.2cm}\\
\rho'\rho(A)v=vA            \ \ ; \ \ A\in\Mg_{r(\rho)}
\end{array}
$$
From this we obtain two faithful conditional
expectations, namely
$$
\begin{array}{l}
\epsilon_\rho=\rho(v^*\rho'(\cdot)v)\in
C(\Mg_{s(\rho)},\rho(\Mg_{r(\rho)}))
\vspace{0.2cm}\\
\epsilon_{\rho'}=\rho'(\bar v^*\rho(\cdot)\bar v)\in
C(\Mg_{r(\rho)},\rho'(\Mg_{s(\rho)})) \ \ \ .
\end{array}
$$
We get also a conditional expectation by
$$
\epsilon^1_\rho=\bar v^*\rho\rho'(\cdot)\bar v \in
C(\Mg_{s(\rho),1},\Mg_{s(\rho)})
$$
and obtain the equation:
$$
\epsilon_{\rho'}\circ\rho'=\rho'\circ\epsilon^1_\rho
$$
Since both $\rho'$ and $\epsilon_{\rho'}$ are faithful, we
conclude that $\epsilon^1_\rho$ is faithful.
Thus we have faithful conditional expectations
$$
\begin{array}{l}
\epsilon_\rho\in C(\Mg_{s(\rho)},\rho(\Mg_{r(\rho)}))
\vspace{0.2cm}\\
\epsilon_\rho'= j_{s(\rho)}\circ\epsilon^1_\rho\circ j_{s(\rho)}
\in C(\rho(\Mg_{r(\rho)})',\Mg_{s(\rho)}')
\end{array}
$$
and hence the index of $\rho$ is finite.
It remains to be proven that $\theta'=j_0\theta$.
One can prove this statement with the same methods used in
\cite{Lo2} (Theorem 4.1).
We sketch here only the main ideas.
If we consider any representative $\bar\rho$ of $j_0\theta$, then
by construction of the $C_0$-conjugate,
$\gamma_\rho=\rho\bar\rho$ is the canonical endomorphism
mapping $\Mg_{s(\rho)}$ into $\rho(\Mg_{r(\rho)})$.
Now let $u$ be a unitary operator implementing
$\rho\rho'$ --- $\rho'$ is any representative of $\theta'$ ---
on $\Mg_{s(\rho)}$. Furthermore, let $v$ be the isometry
which intertwines $\rho\rho'$ and $\id_{s(\rho)}$.
If we follow the argumentation of \cite{Lo2}, we can conclude
that $v_0:=u^*\rho(v)u$ is contained in $\rho(\Mg_{r(\rho)})'$
and we can find a unitary $z\in\rho(\Mg_{r(\rho)})'$ such that
$$
\Gamma:=z^*v_0J_{s(\rho)}v_0^*zJ_{s(\rho)}
$$
implements the canonical endomorphism
$\gamma_\rho=\rho\bar\rho$.
Thus we have $\theta'=j_0\theta$.
\eprf

\bprf (Theorem 3.2)
{\it iii} $\ra$ {\it ii}:
If we suppose that statement {\it iii} is true, then we get:
$$
j_0\theta=j_+\theta
$$
Since $j_0$ is involutive, we obtain together with
Theorem 3.1:
$$
j_0j_+j_0\theta=j_-\theta=j_+\theta
$$

{\it ii} $\ra$ {\it i}:
Suppose now that statement {\it ii} is true.
For each $\theta\in\sec_1(q)$, we obtain for
$\bar\theta:=j_+\theta=j_-\theta$, by applying 
Theorem 2.2,
$$
\theta\bar\theta>i_{s(\theta)} \ \ \ {\rm and} \ \ \
\bar\theta\theta>i_{r(\theta)}  
$$
and Lemma 3.1 implies that $d(\theta)<\infty$.

{\it i} $\ra$ {\it iii}:
Suppose the dimension of $\theta$ is finite, then there exists an
inner massive one-soliton sector $\bar\theta$, such that
$$
\theta\bar\theta>i_{s(\theta)} \ \ \ {\rm and} \ \ \
\bar\theta\theta>i_{r(\theta)}  \ \ \ .
$$
By Theorem 2.2.({\it ii}) we have $\theta j_-\theta> i_{s(\theta)}$.
Let $\bar\rho$ a representative of $\bar\theta$, 
$\bar\rho_-$ a representative of $j_-\theta$ and 
$\rho$ a representative of $\theta$. 
We choose $\rho,\bar\rho,\bar\rho_-$ to be localized in $W_+$.
Since the index of the inclusion 
$\rho(\Ag(W_+)_{r(\theta)})\subset\Ag(W_+)_{s(\theta)}$ is finite,
both $\gamma:=\rho\bar\rho$ and $\gamma_-:=\rho\bar\rho_-$
are inner unitarily equivalent to the canonical 
endomorphism $\gamma_\rho=\rho\bar\rho_0$ \cite{Lo1,Lo2}.
Thus we conclude $\bar\theta=j_-\theta$ and {\it iii}
follows.
\eprf
%From Theorem 2.2 we conclude that $\bar\theta=j_p\theta$, hence
%we have $\bar\theta=j_0\theta$ and {\it iii}
%follows.
%\eprf

Furthermore applying the methods used in \cite{Lo2} 
to the soliton case, one obtains that the dimension function is 
invariant under the charge conjugation.
\blem
Let $\theta\in\sec(q)$ be an inner soliton sector. If
$d(\theta)<\infty$ then $d(j_0\theta)<\infty$ and
one has:
$$
d(\theta)=d(j_0\theta)
$$
\elem
\bdef
The set of all inner soliton sectors generated
%%%%%%%%%%%%%%%%%%%%%%%%%%%%%%%%%%%%%%%%%%%%%%%%%%%%%%%%%%%%%%%%%%%%%%%%%%%%%
\footnote{A subset $\sg$ which generates $\sec(q)$ consists of a 
collection of inner sectors such that each element of 
$\sec(q)$ is a subobject of (or equal to) a 
finite direct sum of finite products of 
elements in $\sg$.}
%%%%%%%%%%%%%%%%%%%%%%%%%%%%%%%%%%%%%%%%%%%%%%%%%%%%%%%%%%%%%%%%%%%%%%%%%%%%%
by all massive inner one soliton sectors which 
fulfill one of the
properties of Theorem 3.2, is denoted by $\sec_f(q)$
and called the set of inner soliton sectors with finite
dimension.
\eef

{\it Remark:} 
By Proposoiton 3.1, the dimension map
respects the multiplication and the direct sum
of inner soliton sectors. 
\bcor
Let $\rho\in\shom(q,x)$ be a soliton homomorphism such that
$e(\rho)$ is contained in $\sec_f(q)$.
Then there exists a co-conjugation, called
{\sl charge-conjugation},
$$
j^*:\sec_f(q)\to\sec_f(q)
$$
such that each antisoliton homomorphism $\bar\rho$ 
satisfies
%$$
%e(\bar\rho)=j^* e(\rho)\in\sec_f(q)
%$$
the properties {\it A1} - {\it A3}.
In addition to that, the dimension map is invariant under 
charge-conjugation $j^*$, i.e.
$d= d\circ j^*$.
\ecor
\bprf
We define $j^*:=j_0$ and on the set  
of all massive inner one soliton sectors with finite 
dimension $\sec_{1,f}(q)$ holds the equation   
$$ 
j^*|_{\sec_{1,f}(q)}=j_0|_{\sec_{1,f}(q)}=j_\pm 
$$
Since $\sec_{1,f}(q)$ generates $\sec_f(q)$ and 
$j_0$ is a co-conjugation, we conclude by applying 
Theorem 3.2 and Proposition 3.1 that the statement of the 
corollary is true. 
\eprf
  
%%%%%%%%%%%%%%%%%%%%%%%%%%%%%%%%%%%%%%%%%%%%%%%%%%%%%%%%%%%%%%%%%%%%%%%%%%%%%%%
\section{Concluding Remarks}
%%%%%%%%%%%%%%%%%%%%%%%%%%%%%%%%%%%%%%%%%%%%%%%%%%%%%%%%%%%%%%%%%%%%%%%%%%%%%%
We have seen that an antisoliton sector can be constructed by 
applying the methods of D.Guido and R.Longo \cite{GuiLo} on the one hand, 
or applying the methods used by K.Fredenhagen \cite{Fre3} on the 
other hand.

We have learned that for soliton sectors with 
finite dimension, each of the described constructions lead 
to the same antisoliton sector. However, one cannot
exclude the possibility that there are massive one-soliton 
sectors with infinite dimension or not. 

Indeed, there are examples for 
theories in which sectors of infinite 
dimension (statistics) arise \cite{Fre6}, but
we have to mention that 
these examples are related to models which describe massless 
particles. 

The construction of $C_0$-conjugates is also possible for massless 
theories, whereas 
for the construction of the $C_\pm$-conjugates one uses that there is 
a mass gap in the energy-momentum spectrum of the theory.

To study the peculiar case where the dimension of a soliton sector is infinite,
one has either to construct models which describe massive 
one-particle states belonging to a sector of infinite dimension, 
or one has to find out whether the construction of $C_\pm$-conjugates
is also possible for the massless examples described in \cite{Fre6}.
    
%%%%%%%%%%%%%%%%%%%%%%%%%%%%%%%%%%%%%%%%%%%%%%%%%%%%%%%%%%%%%%%%%%%%%%%%%%%%%%
\subsection*{Acknowledgment}
I am grateful to Prof. K. Fredenhagen for hints and discussion.
He supported this investigation with many ideas. Thanks are also 
due to my colleagues in Hamburg for careful reading.
%%%%%%%%%%%%%%%%%%%%%%%%%%%%%%%%%%%%%%%%%%%%%%%%%%%%%%%%%%%%%%%%%%%%%%%%%%%% 

%%%%%%%%%%%%%%%%%%%%%%%%%%%%%%%%%%%%%%%%%%%%%%%%%%%%%%%%%%%%%%%%%%%%%%%%%%%%
%   06/07/94 408091439  MEMBER NAME  BIBCMP   (TEXT)     M  TEX
%%%%%%%%%%%%%%%%%%%%%%%%%%%%%%%%%%%%%%%%%%%%%%%%%%%%%%%%%%%%%%%%%%%%%%%%%%%%%
%%%%%%%%%%%%%%%%%%%%%%%%%%%%%%%%%%%%%%%%%%%%%%%%%%%%%%%%%%%%%%%%%%%%%%%%%%%%%

\end{document}